\documentclass[3p]{elsarticle}

\usepackage[colorlinks=true]{hyperref}
\usepackage{amsmath,amsthm,amssymb}
\usepackage{mathtext}
\biboptions{sort&compress}
\newcommand{\rot}{\mbox{rot\,}}  \newcommand{\diverg}{\mbox{div\,}}

\journal{arxiv.org}

\begin{document}

\begin{frontmatter}

\title{Advanced quasistatic approximation}

\author[1,2]{P.V. Tuev\corref{mycorrespondingauthor}}
\ead{p.v.tuev@inp.nsk.su}
\author[1,2]{R.I. Spitsyn}
\author[1,2]{K.V. Lotov}

\address[1]{Budker Institute of Nuclear Physics SB RAS, Novosibirsk, Russia}
\address[2]{Novosibirsk State University, Novosibirsk, Russia}
\cortext[mycorrespondingauthor]{Corresponding author.}

\begin{abstract}
The quasistatic approximation (QSA) is an efficient method of simulating laser- and beam-driven plasma wakefield acceleration, but it becomes imprecise if some plasma particles make long longitudinal excursions in a strongly nonlinear wave, or if waves with non-zero group velocity are present in the plasma, or the plasma density gradients are sharp, or the beam shape changes rapidly. 
We present an extension to QSA that is free from its limitations and retains its main advantages of speed and reduced dimensionality.
The new approach takes into account the exchange of information between adjacent plasma layers. 
We introduce the physical model, describe its numerical implementation, and compare the simulation results with available analytical solutions and other codes.
\end{abstract}


\end{frontmatter}

\section{Introduction}

Particle acceleration in plasma and in particular plasma wakefield acceleration is a promising direction of accelerator development \cite{NJP23-031101}. 
In this method, a driver (a bunch of charged particles or a short laser pulse) propagates through the plasma and creates a Langmuir wave in it, which accelerates the particle bunch called a witness. Both driver and witness move at almost the speed of light, so the plasma-mediated energy exchange between them can be long-lasting and efficient.

The complex phenomena occuring in plasma wakefield accelerators can only be analysed analytically in simplest approximations, which is why numerical simulations are the main method of theoretical research \cite{RAST9-165}. 
Most of the interesting processes take place in a small area of space that moves with the beams. This makes simulations easier, as it allows the use of a short moving window.
However, there are widely different temporal and spatial scales in this problem \cite{NIMA-410-461,PPR31-292}, which can vary by many orders of magnitude, from the wavelength of the laser pulse (about a micron) to the full acceleration length (hundreds of meters) \cite{PRST-AB13-101301,PRST-AB14-091301,PRST-AB15-051301}.
For this reason, it is not always possible to simulate plasma wakefield acceleration with particle-in-cell (PIC) codes based on first-principle equations, and modified (e.g. boosted frame \cite{PRL98-130405,JCP230-5908}) or reduced models have to be used \cite{RAST9-165}.

\begin{figure}[tb]
\centering\includegraphics{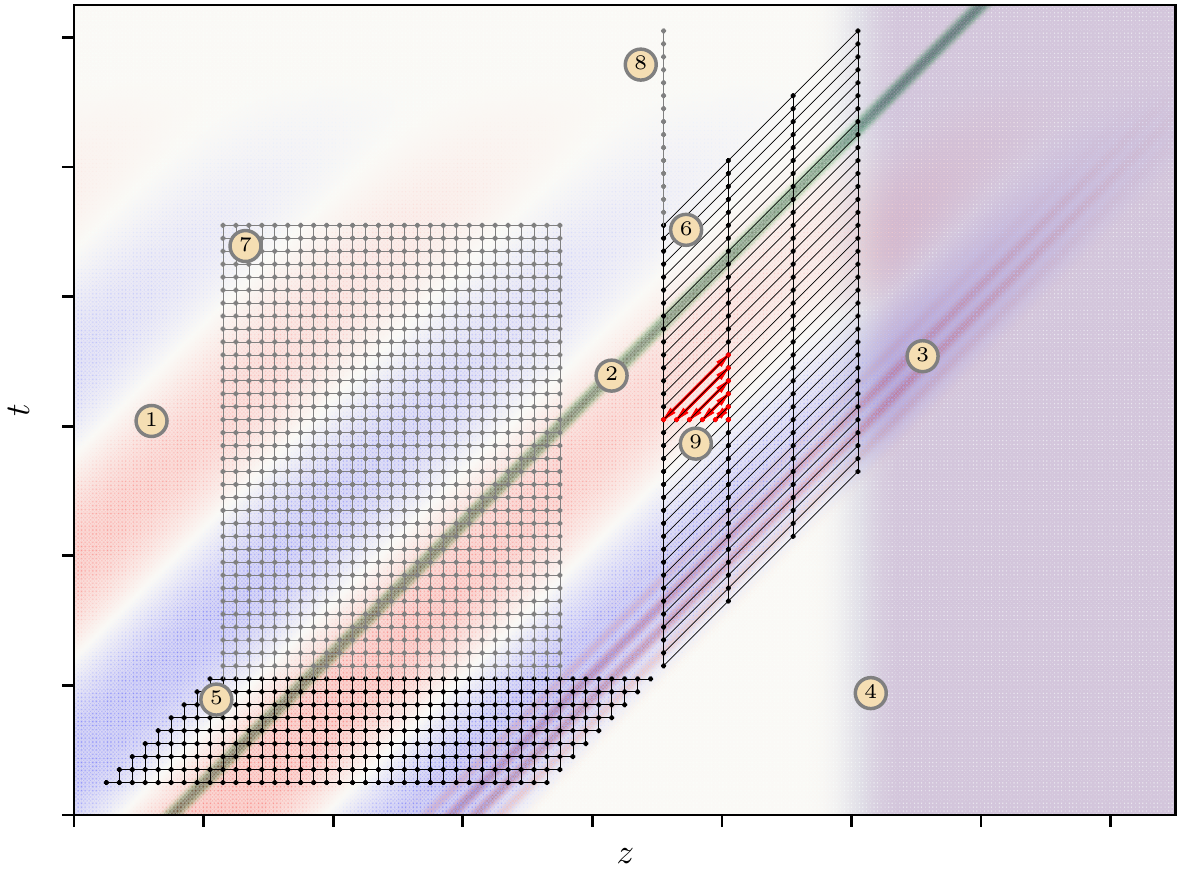}
    \caption{Illustration of objects simulated in the context of plasma wakefield acceleration: plasma wave (1), witness beam (2), laser driver (3), gradient of the plasma density (4). The simulation grids used in conventional PIC codes (5),(7) and in quasistatic codes (6),(8) for modelling beam dynamics (5),(6) and temporal evolution of the plasma wave (7),(8). Duality of interpretation of quasistatic results (9).}\label{fig1-idea}
\end{figure}

One popular reduced model is the so-called quasistatic approximation (QSA) \cite{PRL64-2011, PoP4-217}. 
It has been implemented in a number of codes \cite{PoP22-023103, NIMA-829-350, JCP250-165, PPCF56-084012, PRL121-264801}, sometimes in combination with other simplifications such as envelope equation for the laser pulse \cite{PoP4-217, PoP19-033105, JCP217-658, master} or fluid approximation for the plasma \cite{PRL69-2200,PFB5-2690, PoP5-785}.
The QSA relies on the fact that the properties of investigated objects change much more slowly in the co-moving frame than in the laboratory frame (figure~\ref{fig1-idea}). 
This applies to properties of the plasma wave (feature 1 in figure~\ref{fig1-idea}), densities and velocities of particle beams (feature 2), and intensity of the laser driver (feature 3). 
If the beams propagate in the $z$-direction, the change of variables from the longitudinal coordinate $z$ and time $t$ to 
\begin{equation}\label{e-qs}
    s = z, \qquad \xi = z-ct,
\end{equation}
where $c$ is the speed of light, results in functions that depend on $s$ (at constant $\xi$) much slower than on $\xi$ (at constant $s$) or on $z$ and $ct$ separately.
While conventional PIC codes require small grid steps in both time and longitudinal coordinate to resolve rapid changes in simulated quantities (feature 5), QSA requires small steps only in $\xi$ and allows large steps in $s$ (feature 6). 
This increased grid step makes QSA codes several orders of magnitude faster than conventional PIC codes.
The gain in speed is determined by the ratio of beam evolution length to the plasma or laser wavelength, depending on the driver type. 
Note also that in QSA, the time dependence of simulated quantities at some $z$ can be easily converted into their dependence on $z$ at a fixed $t$ (feature 9), if the beam and plasma properties change insignificantly at the corresponding spatial and temporal intervals.

The QSA is exact if the beams of unchanging shape propagate in a longitudinally uniform plasma. 
In this case, an additional symmetry appears in the problem: identical plasma particles originally located at the same transverse position $\vec r_\perp$, but different $z$, copy the motion of each other with some time delay.
The symmetry leads to another advantage of QSA: the dimensionality of the problem reduces by one when computing the plasma response. 
Namely, one spatial coordinate ($z$) disappears merging with time $t$ into the time-like coordinate $\xi$.
The reduced dimensionality allows a much smaller number of ``macro-particles'' to be used for calculating the plasma response. 
These plasma ``macro-particles'' are not bunches of real particles as in PIC codes, but ``particle jets'' grouping real particles that start their motion from a given transverse position with a given initial momentum, but different $z$.

The third advantage of QSA is its effectiveness in simulating long-term dynamics of the plasma wave.
In PIC codes, to calculate the plasma state at the next point in time, we need information about the current state of the neighbouring plasma layers.
Therefore, it is only possible to simulate the temporal evolution in long regions covering several wakefield periods $\lambda_p$ to minimize the influence of longitudinal boundaries (feature~7 in figure~\ref{fig1-idea}).
In QSA, the state of the neighboring layers is not needed for advancing the plasma in $\xi$ (feature~8).
The model assumes that these layers copy the considered plasma layer with some time delay or advance.
The resulting gain in speed is several times the ratio $\lambda_p / \Delta z$. 
Thanks to this advantage, QSA codes hold the record for simulating long-term evolution of the plasma wave \cite{NatComm11-4753, PPCF64-045003}. 

If the beams change shape slowly, or the unperturbed plasma density is weakly dependent on $z$, then QSA becomes imprecise, but still applicable. 
The inaccuracy arises because of incorrect information exchange between neighbouring $z$-layers. 
The plasma solver assumes that plasma and beam are the same at all $z$, but they are not. 
The beam evolves as it propagates in the plasma, and the initial plasma state may be different at different $z$.
Typically, the region of space the state of which affects the plasma behavior at some $z$ is about the plasma wavelength in the longitudinal direction and does not expand with time, since the group velocity of the Langmuir wave in a cold plasma is zero.
The relative accuracy of QSA in this case can be estimated as the ratio of the plasma wavelength to the typical distance of beam change or the scale of plasma non-uniformity. 
However, if some plasma particles make longer longitudinal excursions in a strongly nonlinear wave, or there are waves with non-zero group velocity in the plasma, or the plasma density gradients are sharp, the QSA becomes less accurate and the error accumulates over time.

In this paper, we present an advanced quasistatic approximation (AQSA), which is free from the limitations of conventional QSA and retains its main advantages of speed and reduced dimensionality.
The new approach takes into account the exchange of information between adjacent plasma layers and therefore allows correct simulation of longitudinal plasma non-uniformities, fast particles appearing in the plasma, and waves with a non-zero group velocity, if these features are resolved by the simulation grid. 
In section~\ref{s2}, we introduce the physical model and discuss its numerical implementation using a two-dimensional geometry as an example.
In section~\ref{s3}, we compare the simulation results with available analytical solutions and with QSA and PIC codes and show that the new model reproduces effects missing in QSA. 
In section~\ref{s4}, we discuss directions of development and the applicability area of the new model.

\section{Extension of the quasi-static approximation}
\label{s2}

\subsection{Physical model}

We start from the Maxwell equations
\begin{gather}
    \label{e1}
    \rot \vec{E} = -\frac{1}{c} \frac{\partial \vec{B}}{\partial t}, \qquad
    \diverg \vec{E} = 4\pi \rho, \\
    \label{e2}
    \rot \vec{B} = \frac{4\pi}{c} \vec{j}  + \frac{1}{c} \frac{\partial \vec{E}}{\partial t}, \qquad
    \diverg \vec{B} = 0, 
\end{gather}
where $\vec{E}$ and $\vec{B}$ are electric and magnetic fields, and $\vec{j}$ and $\rho$ are current and charge densities. The equations can be combined to yield
\begin{equation}\label{m5}
    \Delta \vec{E} - \frac{1}{c^2} \frac{\partial^2 \vec{E}}{\partial^2 t} = 4\pi \nabla \rho + \frac{4\pi}{c^2} \frac{\partial \vec{j}}{\partial t}.
\end{equation}
When passing to the variables $(s, \xi)$, the derivatives change as
\begin{equation}
    \frac{\partial}{\partial z} = \frac{\partial}{\partial s} + \frac{\partial}{\partial \xi}, \qquad
    \frac{\partial}{\partial t} = -c \frac{\partial}{\partial \xi}.
\end{equation}
We assume that all quantities depend slowly on $s$, so the corresponding derivative is a small parameter: $\partial_s  \ll \partial_\xi$. 
In QSA, all small terms containing $\partial_s$ are omitted. 
Here we retain the first-order terms:
\begin{equation}\label{m6}
    \left ( \Delta_\perp +  2 \frac{\partial^2}{\partial s \partial \xi} \right) \vec{E}  
    = 4\pi \hat \nabla \rho - \frac{4\pi}{c} \frac{\partial \vec{j}}{\partial \xi},
\end{equation}
where $\hat \nabla = (\nabla_\perp, \, \partial_s  + \partial_\xi)$, and the subscript $\perp$ denotes two-dimensional (transverse) vectors and operators.
Similarly, for the magnetic field we obtain
\begin{equation}\label{m7}
    \left ( \Delta_\perp +  2 \frac{\partial^2}{\partial s \partial \xi} \right) \vec{B}  
    = - \frac{4\pi}{c} \left[ \hat \nabla \times \vec{j} \right].
\end{equation}
The mixed derivatives in equations (\ref{m6}) and (\ref{m7}) are responsible for propagation of free radiation.

\begin{figure}[tb]
   \centering
   \includegraphics{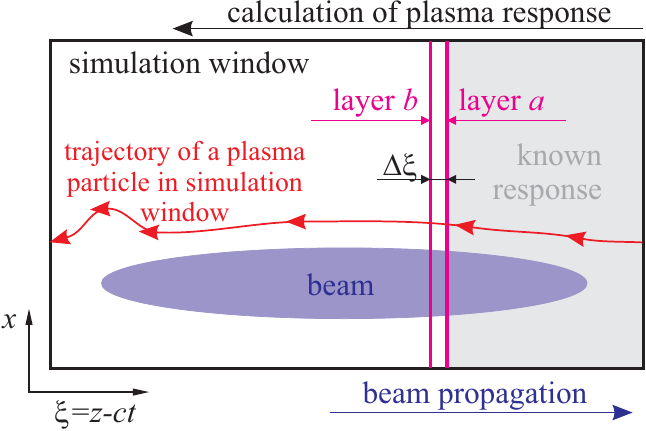}
   \caption{Calculation of plasma response in the quasi-static approximation.}
   \label{fig2-window}
\end{figure}

In addition to the equations for the electromagnetic field, we must also modify the laws of motion of the plasma macro-particles, that is, the laws of changing their parameters.
The parameters include transverse position $\vec{r}_\perp$, three momentum components $\vec{p}$ and charge $q$, and they must be calculated as functions of $\xi$ for a given $s$  (figure~\ref{fig2-window}).
The longitudinal position of a macro-particle is not a parameter, as it is combined with time $t$ and moved from parameters to arguments. 
The mass $m$ is proportional to the charge $q$ and does not need a separate consideration. 
In usual variables, parameters of plasma particles change according to the equations
\begin{equation}\label{e8}
    \frac{d\vec{r}_\perp}{dt} = \vec{v}_\perp, \qquad
    \frac{d\vec{p}}{dt} = q \left( \vec{E} + \frac{1}{c} \left[ \vec{v} \times \vec{B} \right] \right), \qquad
    \frac{dq}{dt} = 0,
\end{equation}
where $\vec{v} = \vec{p} c / \sqrt{p^2 + m^2 c^2}$ is the particle velocity.
Denote an arbitrary parameter by $\chi (s, \xi)$ and the right-hand side of the corresponding equation in (\ref{e8}) by $F$. In the quasistatic variables (\ref{e-qs}), we obtain
\begin{equation}\label{m9}
    \frac{d\chi}{dt} = F = \frac{\partial \chi }{\partial s}\bigg|_\xi \frac{ds}{dt} 
    + \frac{\partial \chi}{\partial \xi}\bigg|_s \frac{d\xi}{dt} 
    = \frac{\partial \chi }{\partial s}\bigg|_\xi v_z 
    + \frac{\partial \chi}{\partial \xi}\bigg|_s (v_z - c),
\end{equation}
whence
\begin{equation}\label{m10}
    \frac{\partial \chi}{\partial \xi}\bigg|_s = \frac{1}{v_z - c} \left( F   - v_z \frac{\partial \chi }{\partial s}\bigg|_\xi \right).  
\end{equation}
Equation (\ref{m10}) always applies to the particle in the ``particle jet'' which at the moment has the longitudinal coordinate $s$.
This particle may initially (before the driver comes) be at some distance from this cross-section.
Thus, equation (\ref{m10}) describes parameters of different physical particles at different $\xi$. 
If we neglect the derivative $\partial_s$, equation (\ref{m10}) reduces to the usual QSA \cite{PRST-AB6-061301}.

To complete the system, equations (\ref{m6}), (\ref{m7}) and  (\ref{m10}) must be supplemented by the equations for charge and current, which are the same as in QSA:
\begin{equation}\label{e11}
    \vec{j} = A \sum_i \frac{q_i \vec{v}_i}{c-v_{z,i}} + \vec{j}_b, \qquad
    \rho = A \sum_i \frac{q_i}{c-v_{z,i}} + \rho_b,
\end{equation}
where $\vec{j}_b$ and $\rho_b$ are current and charge densities of the driver, $A$ is a normalization factor, and summation is over ``particle jets'' crossing the cell in which we calculate $\vec{j}$ and $\rho$. 
The denominators in equation (\ref{e11}) appear because the contribution of a ``particle jet'' to the density and current depends on the macro-particle velocity in the co-moving simulation window. 

The equations for the particle driver are as usual (see, e.g., \cite{PRST-AB6-061301}):
\begin{equation}\label{mb1}
    \frac{d \vec{r}_{b\perp}}{d s} = \frac{\vec{v}_{b\perp}}{c}, \qquad
     \frac{d \xi_b}{d s} = \frac{v_{bz}}{c} - 1, \qquad
     \frac{d \vec{p}_b}{d s} = \frac{q_b}{c} \vec{E} + \frac{q_b}{c^2} \left[ \vec v_b \times \vec B \right], \qquad
     \vec{v}_b = \frac{\vec{p}_b c}{\sqrt{m_b^2 c^2 + p_b^2}},
\end{equation}
where $(\vec{r}_{b\perp}, \xi_b)$ are coordinates of the beam macro-particle in the simulation window, and $m_b$, $q_b$ and $\vec{p}_b$ are its mass, charge and momentum. 
The model can be augmented with the envelope equation for the laser driver and the ponderomotive force acting on the plasma particles \cite{PoP4-217}, but we will not use this option in the paper. 

\subsection{Numerical implementation}

We have added the new features to the existing quasistatic 2D3V code LCODE \cite{lcode}. 
The code can work in Cartesian or axisymmetric geometry. 
We use both geometries for testing the new model, but here we present the equations for the Cartesian geometry only. 
Modifications to the cylindrical geometry are straightforward \cite{manual}. 
In the test cases considered, three field components are identically zero ($E_\varphi = B_r = B_z = 0$ in cylindrical geometry and $E_y = B_x = B_z = 0$ in Cartesian geometry), so we omit the equations for these components.

We calculate the plasma response in LCODE layer-by-layer in the decreasing $\xi$ direction (from right to left in figure~\ref{fig2-window}).
As we need the derivatives of the current with respect to $\xi$ in equation (\ref{m6}) to calculate the field $E_x$, we use the following predictor-corrector scheme:
\begin{itemize}
    \item move the plasma macro-particles from layer $a$ to layer $b$ under the action of the fields of layer $a$;
    \item calculate the current and charge density in layer $b$;
    \item calculate all fields in layer $b$;
    \item move plasma particles from layer $a$ to layer $b$ under the action of the average fields of layers $a$ and $b$;
    \item again calculate the current and charge density in layer $b$;
    \item again calculate all fields in layer $b$;
    \item for the third time, move plasma particles from layer $a$ to layer $b$ under the action of the average fields.
\end{itemize}

\begin{figure}[tb]
\centering\includegraphics[width=8cm]{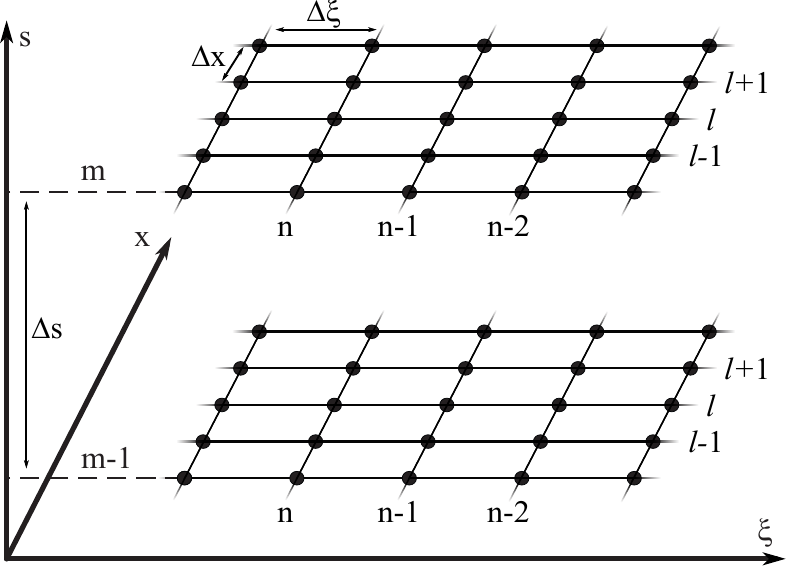}
    \caption{Simulation grid.}\label{fig3-simulation-box}
\end{figure}
To ensure stability of the algorithm, we solve in finite differences, instead of equation (\ref{m6}), the following equation for the transverse electric field:
\begin{equation}\label{m11}
    \left ( \frac{\partial^2}{\partial x^2} +  2 \frac{\partial^2}{\partial s \partial \xi} \right) E_x  - E_x
    = 4\pi \frac{\partial \rho}{\partial x} - \frac{4\pi}{c} \frac{\partial j_x}{\partial \xi} - \tilde E_x,
\end{equation}
where $\tilde E_x$ is some prediction for $E_x$. We present this equation in finite differences as
\begin{multline}\label{m12}
    %\begin{aligned}
     \frac{1}{2} \left( \frac{E_{x, n, l+1}^{m} - 2E_{x, n, l}^{m} + E_{x, n, l-1}^{m}}{\Delta x^2} 
        + \frac{E_{x, n, l+1}^{m-1} - 2E_{x, n, l}^{m-1} + E_{x, n, l-1}^{m-1}}{\Delta x^2} \right) +\\ 
     + \frac{2}{\Delta s} \left( \frac{-3E_{x, n, l}^{m} + 4E_{x, n-1, l}^{m} - E_{x, n-2, l}^{m}}{2\Delta \xi} 
        - \frac{-3E_{x, n, l}^{m-1} + 4E_{x, n-1, l}^{m-1} - E_{x, n-2, l}^{m-1}}{2\Delta \xi}\right) 
        - \frac{E_{x, n,l}^m + E_{x, n,l}^{m-1}}{2} =\\
     = 4\pi \left( \frac{\rho_{n,l+1}^{m-\frac{1}{2}} - \rho_{n,l-1}^{m-\frac{1}{2}}}{2\Delta x} +  
        \frac{1}{c} \frac{j_{x,n,l}^{m-\frac{1}{2}} - j_{x,n-1,l}^{m-\frac{1}{2}} }{\Delta \xi} 
        \right) - \tilde E_{x, n,l}^{m-\frac{1}{2}},
    %\end{aligned}
\end{multline}
where the indexes $l$, $m$, $n$ denote grid layers in $x$, $s$, $\xi$, respectively (figure~\ref{fig3-simulation-box}), $\Delta x$, $\Delta s$, $\Delta \xi$ are grid steps, and half-integer indices denote the half-sum of values at nearby grid points, for example,
\begin{equation}\label{e13}
    \tilde E_{x, n,l}^{m-\frac{1}{2}} = \frac{\tilde E_{x, n,l}^m + \tilde E_{x, n,l}^{m-1}}{2}.
\end{equation}
When we calculate the fields in the layer with index $n$ for the first time (predictor), $\tilde E_{x, n, l}^m = E_{x, n-1, l}^m$, otherwise (corrector) $\tilde E_{x, n, l}^m = E_{x, n-\frac{1}{2}, l}^m$,
with $E_{x, n, l}^m$ found at the predictor step.
In choosing a finite-difference approximation of the field derivatives with respect to $\xi$, we rely on the causality principle.
We assume that no perturbation propagates in the opposite direction to the beams, and that all perturbations in the system are driven by the beams and propagate with them or lag behind them.
In a window moving at the speed of light, this means that subsequent layers in $\xi$ cannot affect the previous ones, so only the right-hand derivatives with respect to $\xi$ should be used. 
This limits the applicability area, but allows for pipeline parallelization of simulations \cite{NIMA-829-350}. 
For the transverse laplacian, we use a stable Crank-Nicolson scheme \cite{MPCPS43-1-50}, which has proven itself in solving the laser envelope equation in plasma wakefield simulations \cite{master, PPCF60-014002}. This results in an implicit numerical scheme which, if necessary, can be generalised to the three-dimensional case using the alternating-direction implicit method \cite{JSIAM3-1-28, JSIAM3-1-42}.

To avoid calculating the derivative with respect to $\xi$ in the longitudinal component of equation (\ref{m6}), we use the continuity equation
\begin{equation}\label{m13}
    \frac{\partial \rho}{\partial t} + \diverg \vec j = 0
\end{equation}
and obtain
\begin{equation}\label{m14}
    \left ( \frac{\partial^2}{\partial x^2} +  2 \frac{\partial^2}{\partial s \partial \xi} \right) E_z
    = \frac{4\pi}{c} \frac{\partial j_x}{\partial x} + 4\pi \frac{\partial }{\partial s} \left( \rho + \frac{j_z}{c} \right),
\end{equation}
which in finite differences yields
\begin{multline}\label{m14fd}
      \frac{E_{z, n, l+1}^{m} - 2E_{z, n, l}^{m} + E_{z, n, l-1}^{m}}{\Delta x^2} +\\
     + \frac{2}{\Delta s} \left( \frac{-3E_{z, n, l}^{m} + 4E_{z, n-1, l}^{m} - E_{z, n-2, l}^{m}}{2\Delta \xi} 
        - \frac{-3E_{z, n, l}^{m-1} + 4E_{z, n-1, l}^{m-1} - E_{z, n-2, l}^{m-1}}{2\Delta \xi}\right) =\\
     = 4\pi \left( \frac{1}{c} \frac{j_{x,n,l+1}^{m} - j_{x,n,l-1}^{m} }{2\Delta x}  +
         \frac{\rho_{n,l}^{m} - \rho_{n,l}^{m-1}}{\Delta s} +
         \frac{1}{c} \frac{j_{z,n,l}^{m} - j_{z,n,l}^{m-1} }{\Delta s} 
        \right).
\end{multline}
Systems of equations (\ref{m12}) and (\ref{m14fd}) are solved using the tridiagonal matrix algorithm.

The equation for the magnetic field can be simplified in two-dimensional geometry. 
Subtracting the $z$-components of the first of equations (\ref{e2}) and the second of equations (\ref{e1}) yields
\begin{equation}\label{m16}
    \frac{\partial }{\partial x} (B_y - E_x) = \frac{4\pi}{c} j_z - 4\pi\rho + \frac{\partial E_z}{\partial s}
\end{equation}
and
\begin{equation}\label{m17}
    \frac{B_{y,n,l}^m - B_{y,n,l-1}^m }{\Delta x} - \frac{E_{x,n,l}^m - E_{x,n,l-1}^m }{\Delta x} =
    \frac{4\pi}{c} j_{z,n,l-\frac{1}{2}}^m - 4\pi\rho_{z,n,l-\frac{1}{2}}^m + 
    \frac{E_{z,n,l-\frac{1}{2}}^m - E_{z,n,l-\frac{1}{2}}^{m-1}}{\Delta s}.
\end{equation}
This equation allows the magnetic field to be calculated to a constant.
In cylindrical geometry, $B_\varphi = E_r = 0$ on the axis, which defines the constant.
In Cartesian geometry, we need the $x$-component of the first of equations (\ref{e2}):
\begin{equation}\label{m18}
    \frac{\partial }{\partial \xi} (B_y - E_x) = -\frac{4\pi}{c} j_x - \frac{\partial B_y}{\partial s}.
\end{equation}
It is sufficient to find the constant for only one value of the transverse index $l$. 
We can choose this value ($l'$) at the periphery of the simulation window, where the currents and fields are low and $\partial_s$ can be neglected.
In finite differences
\begin{equation}\label{m18fd}
    \frac{B_{y,n,l'}^{m} - B_{y,n-1,l'}^{m}}{\Delta \xi} - \frac{E_{x,n,l'}^{m} - E_{x,n-1,l'}^{m}}{\Delta \xi} = 
    \frac{4\pi}{c} j_{x,n-\frac{1}{2},l'}^m,
\end{equation}
whence we find $B_{y,n,l'}^{m}$, knowing the other terms.

To calculate the particle parameters, we use an implicit scheme, which for equation (\ref{m10}) is
\begin{equation}\label{m19}
    \frac{\chi_n^m - \chi_{n-1}^{m}}{\Delta \xi} = \frac{1}{c - v_{z,n-\frac{1}{2}}^m}\left( 
    F_{n-\frac{1}{2}}^m - v_{z,n-\frac{1}{2}}^{m-\frac{1}{2}} \frac{\chi_{n}^{m} - \chi_{n}^{m-1}}{\Delta s}
    \right).
\end{equation}

The boundary conditions have not changed from the documented version of LCODE \cite{manual}.
The transverse boundaries are perfectly conducting walls that reflect particles but absorb their energy (change the particle momentum to some small value). 
There are no fields and no plasma particle motion at the right (front) boundary. 
Boundary conditions at the left (rear) boundary are not needed in quasistatic codes.

The beam solver is also unchanged. 
The equations of motion are solved with the modified Euler method (midpoint method). 
The fields acting on the macro-particle are linearly interpolated to the predicted macro-particle location at half the time step.

\section{Benchmarking the new model}
\label{s3}

\subsection{Laser pulse propagation in plasma}

Let us consider the propagation of a short two-dimensional laser pulse through an underdense plasma of density $n_0$ in the linear regime. This test shows how perturbations with a nonzero group velocity are described by the new model.

The group velocity $v_g$ of the pulse differs from the speed of light $c$ not only because of the plasma, but also because of a finite transverse size of the pulse. 
Adapting the theory \cite{PRE59-1082} to the flat two-dimensional case, we obtain for the group velocity measured on the axis near the focus point
\begin{equation}\label{m20}
    \frac{c - v_g}{c} = \frac{1}{x_0^2 k_0^2} + \frac{k_p^2}{2 k_0^2},
\end{equation}
where $x_0$ is the pulse width at the waist, $k_0$ is the pulse wave number, $k_p = \sqrt{4\pi n_0 e^2/(m_e c^2)}$ is the plasma wave number, $e$ is the elementary charge, and $m_e$ is the electron mass.
The transverse size of the laser pulse $x_s$ grows according to the paraxial theory \cite{PRE59-1082}:
\begin{equation}\label{m21}
    x_s = x_0 \sqrt{1 + \frac{s^2}{s_R^2}},
\end{equation}
where $s_R=k_0 x_0^2/2$ is the Rayleigh length, and the distance $s$ is measured from the waist.

\begin{figure}[tb]
    \centering\includegraphics{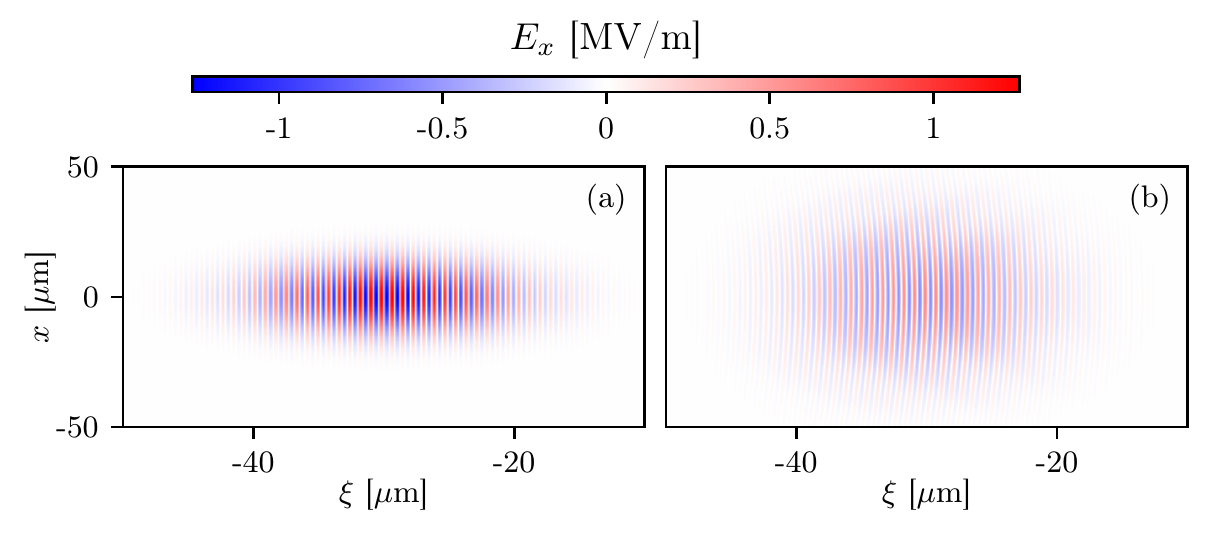}
    \caption{The simulated laser pulse at $s=0$ (a) and $s=2s_R$ (b). 
    For better visibility, only part of the simulation window is shown with different vertical and horizontal scales.
    }\label{fig4-laser}
\end{figure}

Assume the laser pulse is linearly polarized, and its electric and magnetic fields at $s=0$ are
\begin{equation}\label{m22}
    E_x = B_y = E_0 \sin {\left( k_0 \xi \right)} \exp \left( -\frac{x^2}{x_0^2} -\frac{(\xi-\xi_c)^2}{L^2} \right),
\end{equation}
with $\lambda_0 = 2\pi/k_0 = 810$\,nm, $x_0 = 13\,\mu$m, $L = 9\,\mu$m, $E_x = 1.26$\,MV/m (figure~\ref{fig4-laser}). 
The maximum field intensity is initially at $\xi_c = -30\,\mu$m, which is the center of the simulation window $(x, \xi)$ of the size $180\,\mu\text{m}\times 60\,\mu$m. 
The grid steps are $\Delta \xi = \lambda_0/40$, $\Delta x = \lambda_0/4$, $\Delta s = 25\lambda_0$.

\begin{figure}[tb]
    \centering\includegraphics{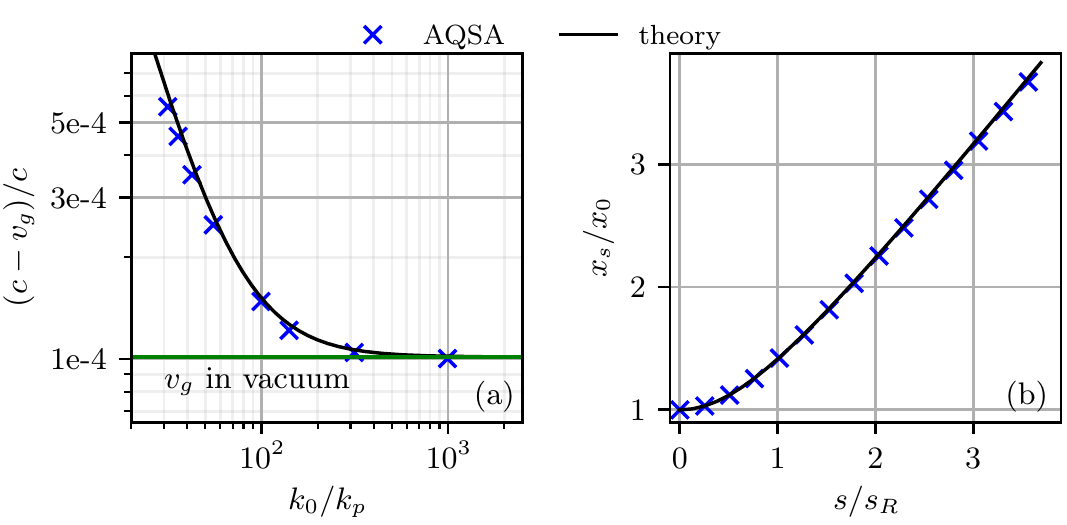}
    \caption{(a) The group velocity of the laser pulse in plasmas of different densities, and (b) growth of the pulse radius in vacuum.}\label{fig5-group_velocity} 
\end{figure}

Unlike other QSA codes \cite{PoP4-217,JCP217-658, master}, the propagation of this pulse is provided by the plasma solver. 
The simulation results quantitatively reproduce the theoretically predicted velocity of the wave packet and its transverse dispersion (figure~\ref{fig5-group_velocity}). 
To find the group velocity from the simulation output, we approximate the pulse envelope on the propagation axis by a Gaussian function and measure the average velocity of the maximum over the time period of $0.13 s_R/c$, starting from the focus point.
The pulse width is measured according to the formula
\begin{equation}\label{m23}
    x_s = 2\sqrt{ \frac{\left< x^2 E_x^2 \right>}{\left< E_x^2 \right>}},
\end{equation}
where angle brackets denote averaging over the simulation window. 
Note that the longitudinal grid step $\Delta s \gg \lambda_0$, so AQSA retains the main advantage of QSA.

\subsection{Plasma with longitudinal density gradient}

\begin{figure}[tb]
    \centering\includegraphics{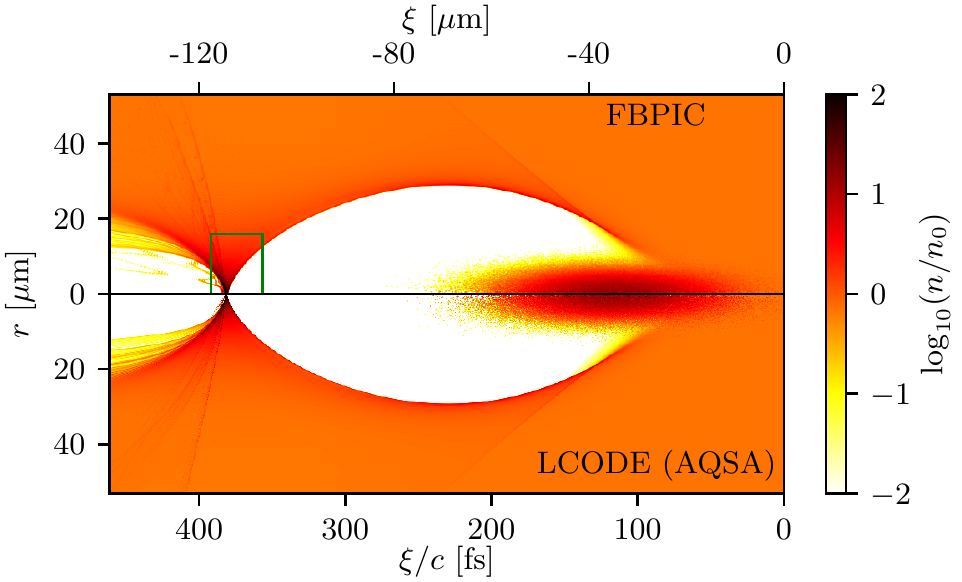}
    \caption{The total electron density $n$ of the beam and plasma for the case when the electron beam propagates in the uniform plasma (at $s=20\,\mu$m) upstream of the density gradient region.
    The green rectangle shows the area detailed in figure~\ref{fig7-density_gradient}. Simulations are performed by FBPIC (upper half) and AQSA model based on LCODE (lower half).}
    \label{fig6-map_density}
\end{figure}

This test shows how the new model describes a strongly nonlinear plasma wave in the presence of density gradients. 
The wave is driven up to the bubble regime by a short axisymmetric electron beam (figure~\ref{fig6-map_density}). 
Plasma electrons at the rear part of the bubble acquire a high longitudinal velocity (close to $c$), travel far forward from their original location and can transfer information from one plasma layer to another.
In a longitudinally uniform plasma, this effect is negated by the same  plasma properties in different cross sections, but in the presence of a density gradient it can be significant. 

\begin{figure}[!t]
    \centering\includegraphics{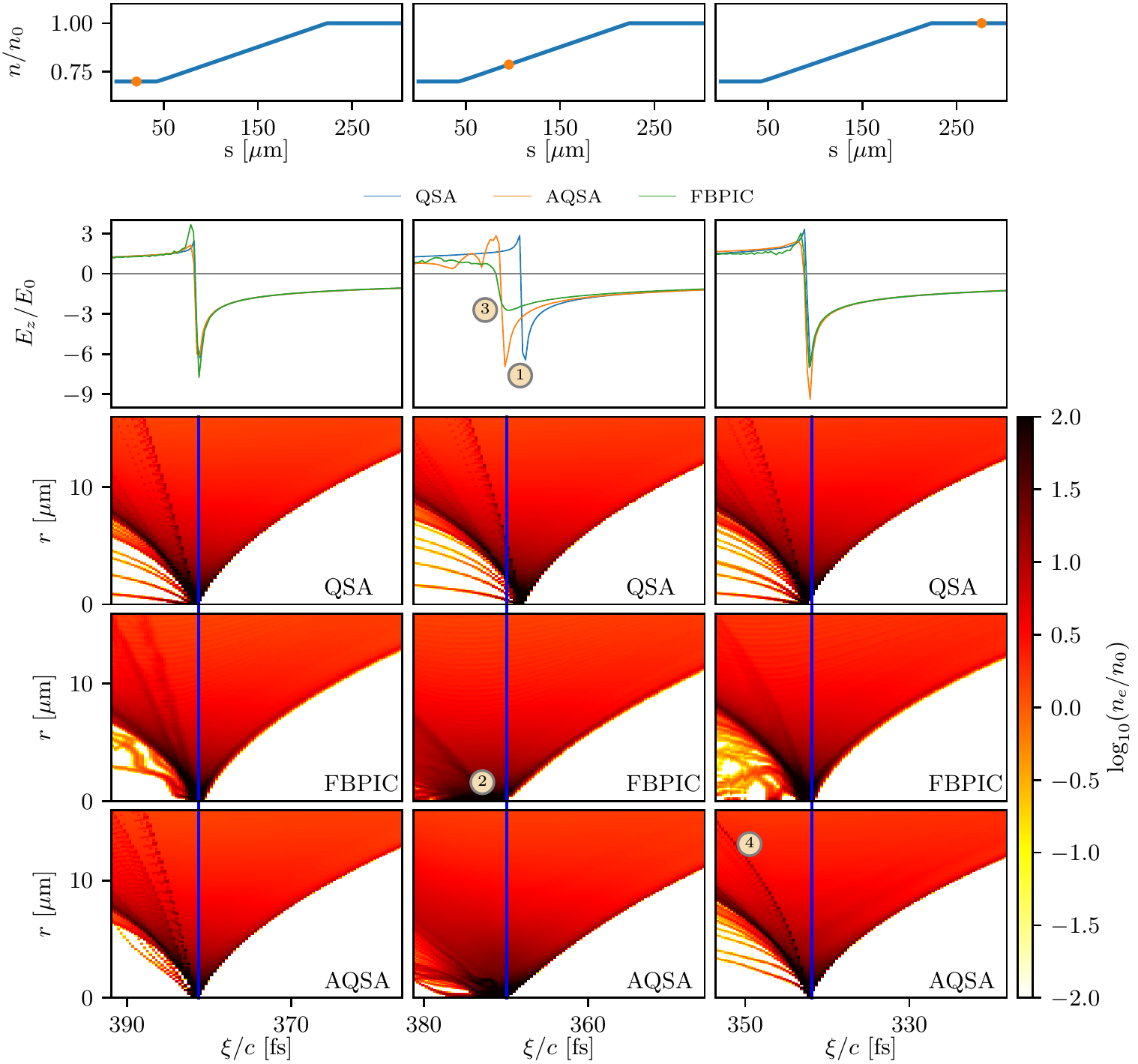}
    \caption{Comparison of QSA, FBPIC, and AQSA simulations of the plasma bubble in uniform plasmas (left and right columns) and in the plasma with the density gradient (middle column). 
    The first (top) row shows the location of the considered cross-sections (orange points) on the plasma density profile. 
    The second row shows the longitudinal electric field $E_z(\xi)$ calculated with the discussed models, and the other rows are the electron density maps obtained with QSA, FBPIC, and AQSA, respectively. 
    The final plasma density $n_0 = 2.5\times 10^{17}\,\text{cm}^{-3}$ defines the electric field unit $E_0 = \sqrt{4 \pi n_0 m_e c^2}$. 
    The blue vertical lines on the density maps show the longitudinal position of the field minimum in AQSA simulations to facilitate comparison. 
    The numbers in circles indicate the features discussed in the text.}
    \label{fig7-density_gradient}
\end{figure}

The electron beam parameters are: root mean square (rms) transverse and longitudinal dimensions $3\,\mu$m and $10\,\mu$m, respectively, relativistic factor $4 \times 10^4$, and the peak beam current 11\,kA.
This beam creates a bubble in the plasma, which is well suited to demonstrate the capabilities of the new model and at the same time has no features with too small spatial scale that are difficult to resolve. 
The longitudinal profile of the plasma density contains a region of linear growth from $1.75\times 10^{17}\,\text{cm}^{-3}$ at $s=50\,\mu$m to $2.5\times 10^{17}\,\text{cm}^{-3}$ at $s=220\,\mu$m, which corresponds to variation of the plasma skin depth $k_p^{-1}$ from $12.7\,\mu$m to $10.6\,\mu$m. 
The density is constant before and after this region (figure~\ref{fig7-density_gradient}, top row). 
The simulations are carried out in the cylindrical coordinates $(r, \xi)$ or $(r, z)$. 

We compare simulations performed with QSA code LCODE, the axisymmetric PIC code FBPIC \cite{CPC203-66}, and the AQSA solver implemented on the basis of LCODE. 
The size of the simulation window is $250\,\mu\text{m}$ in $r$ and $170\,\mu$m in $\xi$. 
In QSA and AQSA simulations, the grid steps are $\Delta r = \Delta \xi = 0.1\,\mu$m, $\Delta s = 10\,\mu$m. 
In FBPIC simulations, $\Delta r = \Delta z = \Delta (ct) = 0.1\,\mu$m. 
To compare quasistatic and PIC simulations, we must present the results in the same coordinates.
Let these be the quasistatic coordinates $(r, \xi)$.
The dependencies on $\xi$ have the meaning of time dependencies at certain $z$.
In a uniform plasma, they can also be interpreted as beam portraits at some time, but not in plasmas with longitudinal density gradients. 
The standard diagnostics in FBPIC do not allow us to output the time dependencies at a fixed $z$, so we save the plasma state and fields every fifth time step (to simplify work with large amounts of data), if the coordinate of interest $s_0$ is in the simulation window. 
Then, for each saved state, we take the values at the five cross-sections nearest to $s_0$ and consider them as five consecutive points of the sought time dependence at $z=s_0$.

All three approaches are consistent in the regions of uniform plasma density (figures~\ref{fig6-map_density} and~\ref{fig7-density_gradient}, left and right columns). 
In the region of density gradient, the classical QSA produces a different plasma response pattern than AQSA and PIC (middle column in figure~\ref{fig7-density_gradient}). 
Unlike QSA, AQSA reproduces the bubble elongation (feature 1 in figure~\ref{fig7-density_gradient}) and the appearance of an extended high-density area near the axis (feature 2) observed in PIC simulations. 
Both effects are caused by longitudinally moving plasma electrons that come from upstream regions where the plasma density is lower and the bubble is longer. 
The reason for the different shape of the field peak in AQSA and FBPIC (feature~3) is not yet completely clear. 
The peak amplitude is determined by the spread of electrons in transverse velocity \cite{PRST-AB6-061301}, so this difference may appear because of the different numerical heating in the two codes. 
The shape of the tail wave \cite{PoP23-103112} observed after the density gradient region (feature~4) is also different in QSA and AQSA or FBPIC. 
This wave (seen as a density ridge) contains high energy plasma electrons accelerated in upstream regions. 
Since the plasma density is lower there, these electrons appear at larger $|\xi|$, and the tail wave front bends, as seen in AQSA and FBPIC simulations, but not in QSA.

To conclude, the AQSA model reproduces the features arising from the longitudinal plasma nonuniformity similarly to the PIC code FBPIC, but 80 times faster, which is close to the theoretical estimate $\Delta s / \Delta \xi = 100$.

\section{Discussion}
\label{s4}

The proposed modification of the quasi-static approximation allows us to extend its applicability area. It opens up the possibility of time-efficient investigation of the laser-plasma interaction, taking into account the physics omitted in the envelope model. The effect of plasma gradients on accelerated particles can be studied more accurately than in the standard QSA. The results obtained are in agreement with both analytical theory and first-principle simulations.

The described model is only the first step towards a more complete and time-efficient simulations of laser and plasma wakefield acceleration based on the quasistatic approximation.
Obviously, the next logical step should be to include plasma particle trapping in the model.
Trapping plasma particles by the wave followed by their acceleration is a widely used method of injecting particles into the wave \cite{APB74-355,PoP19-055501, RMP81-1229}.
The present model does not yet include this process. 
Technically, this is because particle trapping is accompanied by a reduction in the spatial scale of particle parameter change to a level that cannot be resolved.
It is for this reason that we have tested the new model with positive density gradients. 
Negative gradients can lead to particle trapping. 
However, the particles to be trapped can be treated in a special way, similarly to beam particles. 
This approach yields qualitatively correct results with the conventional QSA \cite{PoP17-063106,EPS21-P22004}, and should work even better with AQSA.

The chosen method of numerically solving the equations, although quite reliable, is unlikely the best.
Therefore, another direction for improving the model is to optimise the numerical algorithm.

%In the AQSA model, it is easy to account for field ionisation and other effects that lead to changes in plasma density as the beams pass by. For this, additional terms should be added to the right-hand side of equation (\ref{m10}). So far this has not been possible in quasistatic codes, and the only model of this kind that adds impact ionisation to a QSA code is implemented as an external script \cite{NatComm11-4753}.

While QSA is very efficient in simulating long-term evolution of the plasma wave, it is not certain that AQSA will perform as well on large timescales.
The difference in trajectories and parameters of individual plasma particles at different $s$ increases with the simulation window length that determines the duration of the wave evolution. 
In order to resolve this difference correctly, the grid step $\Delta s$ must be reduced, which depreciates the main advantage of quasistatics.

\section*{Acknowledgement}

The authors are grateful to I.A. Shalimova and A.A. Gorn for helpful discussions. 

%\section*{References}

\end{document}